\documentclass[12pt]{article}
\usepackage[dvipsnames]{xcolor}
\usepackage{amssymb,amsfonts,amsmath,amsthm,amscd,dsfont}

\usepackage{fullpage}
\usepackage[colorlinks]{hyperref}
\usepackage[nameinlink,capitalise]{cleveref}
\usepackage{enumerate}  

\hypersetup{
  linkcolor=[rgb]{0,0,0.4},
  citecolor=[rgb]{0, 0.4, 0},
  urlcolor=[rgb]{0.6, 0, 0}
}

\usepackage{cleveref}

\newcommand{\F}{\mathbb{F}}
\newcommand{\N}{\mathbb{N}}
\newcommand{\Z}{\mathbb{Z}}
\newcommand{\Q}{\mathbb{Q}}
\newcommand{\RR}{\mathbb{R}}

\newcommand{\vx}{\mathbf{x}}
\newcommand{\va}{\mathbf{a}}
\newcommand{\vy}{\mathbf{y}}

\newcommand{\set}[1]{\left\{ #1 \right\}}

\newcommand{\alg}[1]{{\left\Vert #1 \right\Vert}_A}
\newcommand{\Ltwo}[1]{{\left\Vert #1 \right\Vert}_2}

\newcommand{\Det}{\mathsf{Det}}

\DeclareMathOperator{\rank}{\mathsf{rank}}

\newtheorem{theorem}{Theorem}[section]
\newtheorem{definition}[theorem]{Definition}
\newtheorem{conjecture}[theorem]{Conjecture}
\newtheorem{claim}[theorem]{Claim}
\newtheorem{lemma}[theorem]{Lemma}
\newtheorem{corollary}[theorem]{Corollary}

\title{Möbius-Pseudorandomness for Algebraic Circuits}
\author{Gil Kalai\thanks{Efi Arazi School of Computer Science, Reichman University and Einstein Institute of Mathematics, Hebrew University of Jerusalem. Email: \texttt{gil.kalai@gmail.com}. Supported by ERC grant 834735 and by an ISF grant 2669/21.}
\and
Mrinal Kumar\thanks{Tata Institute of Fundamental Research, Mumbai, India. Email: \texttt{mrinal@tifr.res.in}.  Research supported by the Department of Atomic Energy, Government of India, under project number RTI400112, and in part by Google and Premji Invest Research Grants.}
\and
 Ben Lee Volk\thanks{Efi Arazi School of Computer Science, Reichman University, Israel. Email: \texttt{benleevolk@gmail.com}. The research leading to these results has received funding from the Israel Science Foundation (grant number 843/23). }}

\date{}

\begin{document}

\maketitle

\abstract{
We define a notion of Möbius-pseudorandomness for integer sequences generated by multivariate polynomials computed by small and simple algebraic circuits. This mirrors a related definition of Kalai that concerns boolean functions computed by small boolean circuits.

We then prove Möbius-pseudorandomness results for polynomials computed by algebraic circuits with few multiplication gates (of any depth), for polynomials with small-dimensional partial derivative space, and for the determinant and permanent.
We conjecture that these results hold in general for polynomial-size algebraic circuits.
}

\section{Introduction}
\label[section]{sec:intro}

\subsection{Möbius-Pseudorandomness}

Recall that the Möbius function $\mu : \N \to \set{-1,0,1}$ is defined as follows: $\mu(m)=0$ if $m$ isn't square-free, and $\mu(m)=(-1)^k$ if $m$ is square-free and the number of its prime factors is $k$.

A sequence $\set{a_n}_{n \in \N}$ is said to be \emph{Möbius-pseudorandom} if it has vanishingly small correlation with the Möbius function. More formally:

\begin{definition}
\label[definition]{def:mob-pseudorandom}
Let $\set{a_n}_{n \in \N}$ be a bounded integer sequence (or more generally, a sequence of complex numbers). We say that it is \emph{Möbius-pseudorandom} if
\begin{equation}
\label{eq:mobius-pseudorandom}
\frac{1}{N} \left| \sum_{m=0}^{N-1} \mu(m)a_m \right| = o(1).
\end{equation}
\end{definition}

In what follows, it is slightly more convenient to sum over integers from $0$ to $N-1$ and thus we may define arbitrarily $\mu(0)=0$, as this clearly does not matter with respect to the pseudorandomness property.

Similarly, one can define a function to be \emph{Liouville-pseudorandom} if the same holds with the Liouville function $\lambda(m)$ instead of $\mu(m)$. The Liouville function $\lambda : \N \to \set{-1,1}$ is defined as $\lambda(m) = (-1)^{k}$ where $k$ is the number of prime factors of $m$, counted with multiplicity. All the results and conjectures in this paper hold with respect to Liouville-pseudorandomness as well.

This notion of Möbius-pseudorandomness is well studied in number theory and in other areas,\footnote{It is often called there \emph{Möbius-randomness}. However, borrowing the standard terminology of computational complexity theory, we prefer to use ``pseudorandomness'' rather than ``randomness'' as neither the Möbius function nor the sequence we consider are truly random; rather, we investigate the amount to which such sequences correlate with the Möbius function.} and a significant amount of work was devoted to proving that certain sequences are Möbius-pseudorandom. Sarnak \cite{Sarnak2012} has made the following conjecture:

\sloppy
\begin{conjecture}[Sarnak's conjecture] Every deterministic bounded sequence is Möbius-pseudorandom.
\label[conjecture]{c:sarnak}
\end {conjecture} 

The notion of a ``deterministic sequence'' arises from topological dynamics. For Boolean sequences $f(n), n \in \mathbb N$, the definition is especially simple: $f$ is ``deterministic'' if the number of patterns
\[
(f(n+1),f(n+2),\dots , f(n+m)):n \in \mathbb N
\]
is $\exp (o(m))$. See \cite{Sarnak2012} for the general definition. (For the case of general bounded functions $f$ the condition is that for every real $\epsilon >0$ all these patterns can be covered by $\exp (o(m))$ balls of radius $\epsilon$ in the $\ell_\infty$ metric.) 

There is a variety of classes of sequences for which  
Möbius-pseudorandomness holds, including various special cases of Sarnak's \cref {c:sarnak}. For example, the following classical theorem goes back to Davenport.

\begin {theorem}[\cite{Davenport37, Hua65}]
\label[theorem]{thm:davenport}
\begin{enumerate}[(i)]
\item For an irrational number $\alpha$, the sequence $f(n)=e^{\pi i \alpha n}$ is Möbius-pseudorandom. 
\item  For a polynomial $p(n)$ the sequence $f(n)=e^{\pi i \alpha p(n)}$ is Möbius-pseudorandom. 
\end{enumerate}
\end {theorem}

A far-reaching extension of Davenport's theorem was proved by Green and Tao \cite {Green-Tao12}. For another important special case of Sarnak's conjecture see Bourgain, Sarnak, and Ziegler \cite {BSZ12} and the surveys \cite {FKPL18,Ziegler2026}. A general result about \emph{automatic sequences} was proved by M\"ullner \cite {Mullner-2017}.

\subsection {Computational complexity and M\"obius pseudorandomness}
\label[section]{sec:intro:cc-mob}

Sarnak \cite {Sarnak2010,Sarnak2012} raised the question of whether every function that can be computed in polynomial time is Möbius-pseudorandom. Here, in order to make sense of the question one needs to fix an encoding of integers. It is natural to consider integers encoded using binary digits, and therefore the assumption that the function is computed in polynomial time means that there exists a polynomial time algorithm that given the binary digits of $m$ computes $f(m)$, in time which is polynomial in the number of binary digits of $m$. To that end, we introduce some notation. For an integer $m < N=2^n$ let $(m_1, \ldots ,m_n)$ denote its binary digits.

\begin{definition}[Möbius-pseudorandomness of Boolean Functions]
\label[definition]{def:mob-pr-boolean}
Let $f=\set{f_n}_{n \in \N}$ be a sequence of $n$-variate boolean functions. We say $f$ is \emph{Möbius-pseudorandom} if
\[
\frac{1}{N} \left| \sum_{m=0}^{N-1} \mu(m)\exp(i \pi \cdot f_n(m_1, \ldots, m_n)) \right| = o(1)
\]
where $N=2^n$ and $(m_1, \ldots, m_n)$ are the binary digits of the integer $m$.
\end{definition}

We may then consider the complexity of the sequence of functions $\set{f_n}_{n \in \N}$ where $f_n(x_1, \ldots, x_n)$ outputs, on input $m_1, \ldots, m_n$, the integer $f_n(m)$. $f_n$ is efficiently computable if this computation can be done in time polynomial in $n$. Here one may consider either non-uniform computation (using circuits) or uniform computation (in which there exists a single Turing machine that computes $f_n(m)$ for every $n$ and $m$).

Sarnak, however, expected a negative answer to his conjecture, since he expressed his belief that efficient algorithms for integer factoring, and hence for computing $\mu$ do exist. For completeness, we state here a strong form of a positive answer to Sarnak's question. 

\sloppy
\begin {conjecture} 
\label[conjecture]{conj:non-sarnak}
Every polynomial-time computable function $f$ is Möbius-pseudorandom.
\end {conjecture}

\cref{conj:non-sarnak}, while plausible,  
is even stronger than the assertion that there is no polynomial-time factoring algorithm (which is considerably stronger than the ${\bf P \ne NP}$ conjecture).

Kalai initiated a line of work whose purpose is to study Sarnak's conjecture for various limited complexity classes.
The goal is to offer computational complexity conjectures or theorems that on the one hand use computational classes that can be shown to be limited by current techniques (like bounded depth circuits, although we note that sequences represented by bounded depth circuits need not be \emph{deterministic} in the sense of \cref{c:sarnak}), and on the other hand, may include concrete classes of functions that occurred naturally in number theory. A far-reaching goal would be to find computational complexity setting that is rich enough to include many concrete results about Möbius-pseudorandomness from number theory. It seems that arithmetic formulas and circuits could be especially relevant.

Following Kalai's \cite{Kalai12} suggestion, Möbius-pseudorandomness results were proved for various complexity classes.

Green \cite{Green2012} has proved that boolean functions in $\mathsf{AC}^0$ are Möbius-pseudorandom by estimating the low-degree Fourier-Walsh coefficients of the Möbius function. Bourgain \cite{Bourgain2013, bourgain2013mobius, Bourgain2016} obtained some improved estimates on the Fourier-Walsh coefficients, that also allowed him to prove that \emph{monotone} boolean functions are Möbius-pseudorandom. We give more details on Bourgain's results in \cref{sec:preliminaries}, as his results are very relevant to ours.

Extending these results to larger complexity classes is an open problem. We record two conjectures in that regard: 

\begin {conjecture}[Möbius-pseudorandomness conjecture for $\mathsf{AC}^0(p)$]
\label[conjecture]{conj:acc}
Let $p$ be a fixed prime.
Suppose $f = \set{f_n}_{n \in \N}$ is in $\mathsf{AC}^0(p)$, that is, it can be computed by a polynomial size bounded depth Boolean circuit with $\bmod p$ gates. Then $f$ is Möbius-pseudorandom.
\end {conjecture}

\begin {conjecture}[Möbius-pseudorandomness conjecture for low degree polynomials]
\label[conjecture]{conj:ldp}
Suppose $f$ can be represented by a polynomial over $\mathbb{Z}/p\mathbb{Z}$ of degree at most $\mathrm{polylog}(n)$: that is, there exists an $n$-variate polynomial $P$ over $\mathbb{Z}/p\mathbb{Z}$, of degree at most $\mathrm{polylog}(n)$, such that for every $x \in \set{0,1}^n$, $f(x)=P(x)$. Then $f$ is Möbius-pseudorandom.
\end {conjecture}

An important theorem by Razborov and Smolensky \cite{Razborov87, Smolensky87, Smolensky93} implies that \cref{conj:ldp} implies \cref{conj:acc}.

We remark that \cref{conj:ldp} is known to hold for \emph{linear} polynomials, by  Bourgain's result on the Fourier-Walsh coefficients mentioned above. A special case of degree-2 polynomials is Möbius-pseudorandomness for the Rudin-Shapiro sequence.
This case was settled by Tao \cite{Tao2012RudinShapiro} who relies on the Katai-Bourgain-Sarnak-Ziegler orthogonality criterion.

One should also note the difference between \cref{conj:ldp}, which refers to low-degree boolean functions in the binary digits of an integer $n$, with degree that can possibly grow with $n$, and \cref{thm:davenport}, that deals with polynomials in $n$ (of fixed degree).

\subsection{Möbius-Pseudorandomness in Algebraic Complexity}

Similar to the boolean case, one could consider sequences that can be encoded using ``simple'' objects in algebraic complexity theory, which studies the computation of polynomials using arithmetic operations. Our goal in this note is to propose some conjectures along these lines, and prove some special cases of them.

\begin{definition}[Möbius-pseudorandomness for polynomials]
\label[definition]{def:mob-polynomial-pseudorandom-eval}
\sloppy
Let $f=\set{f_n}_{n \in \N}$ be a sequence of $n$-variate integer polynomials. We say $f$ is \emph{Möbius-pseudorandom} if
\[
\frac{1}{N} \left| \sum_{m=0}^{N-1} \mu(m)\exp(i \pi \cdot f_n(m_1, \ldots, m_n)) \right| = o(1)
\]
where $N=2^n$ and $(m_1, \ldots, m_n)$ are the binary digits of the integer $m$.
\end{definition}

For readers versed in algebraic complexity theory, our choice to focus on \emph{evaluations} of the polynomial over boolean inputs may seem odd: Since algebraic complexity theorists are usually concerned with polynomials as formal objects (rather than the functions that they compute) it might seem natural to focus on the sequence of \emph{coefficients} of such polynomials. However, despite considerable efforts and multiple iterations, we were unable to find a general enough ``coefficient version'' of \cref{def:mob-polynomial-pseudorandom-eval} that makes sense. A natural attempt would be to look at the coefficient sequence of multilinear polynomials, but then the $\frac{1}{N}$ normalization in \cref{def:mob-polynomial-pseudorandom-eval} would imply that any homogeneous multilinear polynomial will be Möbius-pseudorandom, regardless of its complexity, as most of its coefficients will be $0$. Different normalizations then lead to trivial constructions of such polynomials that correlate well with the Möbius function, especially since algebraic circuits are a non-uniform model of computation.

\subsection{Our Results}

In this paper we prove Möbius-pseudorandomness for certain classes of polynomials computed by small algebraic circuits in various models.

\begin{theorem}
\label[theorem]{thm:intro:main}
The following classes of polynomials are Möbius-pseudorandom:
\begin{itemize}
    \item Polynomials computed by depth-$3$ circuits of top fan-in $o(n)$ (\cref{sec:depth-3})
    \item Polynomials whose first-order partial derivatives span a space of dimension $o(n)$ (\cref{sec:pd})
    \item Polynomials computed by circuits with $o(n)$ multiplication gates (\cref{sec:few-mult}).
\end{itemize}
\end{theorem}

The definitions of the corresponding models appear in the relevant sections.

In addition, in \cref{sec:determinant}, we prove that the determinant and permanent polynomials, the most important polynomials in algebraic complexity, are Möbius-pseudorandom as well.

In the spirit of \cref{conj:non-sarnak}, we conjecture that such results hold in much greater generality than what we are able to prove.

\subsection{Techniques}
\label[section]{sec:techniques}

The main tool we use is a strong correlation bound by Bourgain \cite{bourgain2013mobius} that bounds the correlation of the Möbius function with Fourier-Walsh characters over $\F_2$. As mentioned in \cref{sec:intro:cc-mob}, such a bound can be used to show that boolean functions in $\mathsf{AC}^0$, and monotone boolean functions, are Möbius-pseudorandom. It is perhaps slightly more surprising that this bound also has applications for showing Möbius-pseudorandomness for \emph{algebraic} circuits. However, it turns out that in many settings if $f(\vx)$ is an integer polynomial computed by some limited class of circuits, we are able to give an upper bound on the algebra norm (that is, the $L_1$ norm of the Fourier-Walsh transform) of the boolean function $(-1)^f$. Such bounds, combined with Bourgain's \cite{bourgain2013mobius} correlation upper bound as well as the estimates of \cite{Bourgain2016}, imply that those polynomials are Möbius-pseudorandom.

For the determinant polynomial, the situation is slightly more complicated. However, here we are able to prove that the determinant modulo 2 can be \emph{approximated} (in $L_2$ norm) by a function with small Fourier-Walsh $L_1$ norm, which is sufficient for the argument to go through.

\section{Preliminaries}
\label[section]{sec:preliminaries}
In this section we state some preliminary results that will be used in the next sections.

As mentioned in \cref{sec:techniques}, Bourgain \cite{bourgain2013mobius, Bourgain2013, Bourgain2016} (following Green \cite{Green2012}) has obtained several bounds on the Fourier-Walsh coefficients of the Möbius function. In particular, \cite{bourgain2013mobius} provides a uniform $2^{-n^{1/10}}$ upper bound on the magnitude of each such coefficient. However, considering low-degree and high-degree coefficients separately slightly improves the parameters in our results.

The following lemma obtains an improved estimate for high-degree coefficients. Its proof is implicit in \cite{bourgain2013mobius}. Since the exact statement does not appear in \cite{bourgain2013mobius}, we detail some of the calculations.

\begin{lemma}[Implicit in \cite{bourgain2013mobius}]
\label[lemma]{lem:fourier-correlation-high-degree}
    Let $\chi_S : \F_2^n \to \set{-1,1}$ be a Fourier-Walsh character of $\F_2^n$ corresponding to $S \subseteq [n]$ that is, $\chi_S(x) = (-1)^{\sum_{i \in S}x_i}$. If $|S| \ge \sqrt{n} \log^2 n$, then there exists an absolute constant $\delta > 0$ such that, for sufficiently large $n$,
    \[
    \frac{1}{N} \left| \sum_{m=0}^{N-1} \mu(m) \chi_S(m_1, \ldots, m_n) \right| \le 2^{-\delta |S|^2/n}
    \] 
    where $N=2^n$.
\end{lemma}

\begin{proof}
Fix $S$ with $|S| = t$ and set $h=t^2 / n$.

Following the proof in \cite{bourgain2013mobius} (see also Lemma 1 of \cite{MR10}), we need to bound the so-called Type-I and Type-II sums.

Set $M_0 = 2^{at}$ and $L=2^{bh}$, where $a>0$ is a small constant and $b>0$ is sufficiently small with respect to $a$. For $M\geq M_0$, Bourgain's \cite{bourgain2013mobius} Equation 2.31 upper bounds Type-II sums by
\[
T_{\mathrm{II}} \le N \left( L^{-c_1} + L^2 M^{-c_2} + L^{C} 2 ^{-c_3 \log_2M |S|/n} \right)
\]
where $c_1, c_2, c_3, C$ are absolute constants (with Bourgain's \cite{bourgain2013mobius} parameter $\varepsilon$ fixed).
Using the assumptions on $M$ and $L$,
\[
T_{\mathrm{II}} \le N \left( 2^{-c_1 bh} + 2^{2bh - c_2 at} + 2^{(Cb - c_3a)h}\right),
\]
where in the third summand we used the fact that $h=t^2 / n$. Since $h \le t$, we get that 
\[
T_{\mathrm{II}} \le N \cdot 2^{-\delta h}
\]
for some $\delta > 0$.

The parameter restrictions preceding (2.3) in \cite{bourgain2013mobius} hold: writing $u=\log_2 M$ and $\rho=\log_2 L = bh$, we have that for large enough $n$,
\[
\log^2 n \ll \rho <u/100,
\]
since $h=t^2 / n \ge \log^4 n$ and $u/100 \geq at/100 > bh$, since $h\le t$ and $b$ can be picked sufficiently small with respect to $a$.

For $M \le M_0$, Type-I sums are upper bounded in equations (3.2$'$) and (1.3) of \cite{bourgain2013mobius}, that give the upper bound
\[
T_{\mathrm I}
\le N n^2 M 2^{-c_0 |S|}
\le N n^2\,2^{(a-c_0)t}
\le N n^2\,2^{-c_0 t/2},
\]
if $a<c_0/2$.

The bounds on Type-I and Type-II sums now imply
\[
\left| \sum_{m=0}^{N-1} \mu(m) \chi_S(m_1, \ldots, m_n) \right| \le N \cdot n^{O(1)} \cdot 2^{-\delta' h} \le N\cdot 2^{-\delta h}
\]
for some constants $\delta',\delta > 0$ (since $h \ge \log^4 n$, the polynomial factor can be comfortably absorbed by choosing a slightly smaller $0<\delta<\delta'$).
\end{proof}

We also require the following estimate by Bourgain \cite{Bourgain2016} on the total mass of the low-degree Fourier-Walsh coefficients of the Möbius function.

\begin{theorem}\label[theorem]{thm:bourgain-low-degree}
Let $\set{\hat{\mu}(S) : S \subseteq [n]}$ denote the Fourier-Walsh coefficients of the Möbius function $\mu$ (thought of as a function $\mu : \set{0,1}^n \to \set{-1,0,1}$ whose input is the binary digits of an integer). Then there exists a constant $\rho>0$ such that for every constant $C$,
\[
\sum_{A \subseteq [n], |A|\leq \rho n} \hat{\mu}(A)^2 \le n^{-C}.
\]
\end{theorem}

For a function $f : \F_2^n \to \mathbb{R}$ we denote by $\set{\hat{f}(S)}_{S \subseteq [n]}$ its Fourier-Walsh coefficients, and by 
$\alg{f}$ its \emph{algebra norm} (or $L_1$ norm), $\alg{f} = \sum_{S \subseteq [n]} |\hat{f}(S)|$. It holds that $\alg{f \cdot g} \le \alg{f} \cdot \alg{g}$.

We further define $\Ltwo{f} = \left( \mathbb{E}_{x} [f(x)^2] \right)^{1/2}$ where the expectation is over a uniformly random $x$.

The following lemma gives an upper bound on the correlation of a function with small algebra norm with the Möbius function.

\begin{lemma}
\label[lemma]{lem:corr-algebra-norm}
There exists a universal constant $\varepsilon > 0$ with the following property:
Let $f : \F_2^n \to \mathbb{R}$ be a function with $\Ltwo{f} \le 1$, and suppose that $\alg{f} \le 2^{\varepsilon n}$. Then 
\[
\frac{1}{N} \left| \sum_{m=0}^{N-1} \mu(m) f(m_1, \ldots, m_n) \right| =o(1)
\]
(where $N=2^n$ and $(m_1, \ldots, m_n)$ are the binary digits of the integer $m$).
\end{lemma}

A weaker form of \cref{lem:corr-algebra-norm} in which the assumption $\alg{f} \le 2^{\varepsilon n}$ is replaced by $\alg{f} \le 2^{\varepsilon n^{1/10}}$ can be directly deduced from the main theorem of \cite{bourgain2013mobius}, that gives a uniform upper bound on the correlation of $\mu$ with any Fourier-Walsh character. However, by exploiting \cref{thm:bourgain-low-degree} and the improved correlation upper bound from \cref{lem:fourier-correlation-high-degree}, we are able to obtain the stronger version. Note that by Cauchy-Schwarz, any boolean function has $\alg{f} \le 2^{n/2}$.

\begin{proof}
We think of $\mu$ as a function $\mu : \F_2^n \to \set{-1,0,1}$ by identifying an integer $m \in [0,2^n-1]$ with its binary digits. The Fourier-Walsh coefficients of $f$ and $\mu$ are then denoted $\hat{f}(S), \hat{\mu}(S)$ for $S \subseteq [n]$, respectively. Expanding both functions in the Fourier basis and using the orthonormality of the Fourier characters, we obtain
\begin{align*}
\left| \frac{1}{N}\sum_{m=0}^{N-1} \mu(m_1, \ldots, m_n) f(m_1, \ldots, m_n) \right| &= 
\left| \sum_{S \subseteq [n]} \hat{f}(S) \hat{\mu}(S) \right| \\
& \le 
\left| \sum_{|S| \le \rho n} \hat{f}(S) \hat{\mu}(S) \right| + 
\left| \sum_{|S| > \rho n} \hat{f}(S) \hat{\mu}(S) \right|.
\end{align*}
Here $\rho$ is a small constant to be picked later.
By Cauchy-Schwarz, we can upper bound the right hand side by 
\begin{equation}
\label{eq:ub-cs}
\left( \sum_{|S| \le \rho n} \hat{f}(S)^2 \right)^{1/2} \cdot 
\left( \sum_{|S| \le \rho n} \hat{\mu}(S)^2 \right)^{1/2} + 
\max_{S \subseteq [n], |S| > \rho n} |\hat{\mu}(S)| \cdot 
\left( \sum_{|S| > \rho n} |\hat{f}(S)| \right).
\end{equation}
The assumption $\Ltwo{f} \le 1$ and Parseval's identity, along with \cref{thm:bourgain-low-degree}, imply that the first term in \eqref{eq:ub-cs} is upper bounded by $n^{-C}$ for any constant $C$, assuming $\rho$ is picked to be a small enough constant to apply \cref{thm:bourgain-low-degree}.

\cref{lem:fourier-correlation-high-degree} and the assumption on $\alg{f}$ imply that the second term is at most
\[
2^{-\delta \rho^2 n} \cdot 2^{\varepsilon n}
\]
where $\delta$ is a small constant (depending on $\rho$). Picking $\varepsilon = \delta \rho^2 / 2$ finishes the proof.
\end{proof}

\section{Depth-3 Formulas}
\label[section]{sec:depth-3}

In the same spirit of Kalai's conjecture, and the results of Green \cite{Green2012} and Bourgain \cite{Bourgain2013, Bourgain2016, bourgain2013mobius}, we conjecture that ``simple'' polynomials are Möbius-pseudorandom, where simplicity is measured with respect to low algebraic complexity. Such results were not previously known even for very simple algebraic models.

\begin{definition}
\label[definiton]{def:depth-3-ml-formula}
Let $f \in \Z[\vx]$ be an $n$-variate multilinear polynomial. We say $f$ is computed by a depth-3 formula of top fan-in $s$ over $\Z$ if
\[
f = \sum_{i=1}^s \prod_{j=1}^{d_i} \ell_{i,j} (x_1, \ldots, x_n)
\]
where for every $i,j$, $\ell_{i,j} \in \Z[\vx]$ is a linear function with integral coefficients.
\end{definition}

There are some natural submodels of depth-$3$ formulas. Recall that we assume $f$ is a multilinear polynomial. If, for every $i$, the linear functions $\ell_{i_1}, \ldots, \ell_{i,d_i}$ are variable disjoint, the formula is called a \emph{depth-$3$ multilinear formula}.

One may even further restrict to \emph{set-multilinear formulas}, defined as follows.
Let $\vx=(x_1, \ldots, x_n)$ denote a tuple of $n$ variables. A polynomial $f \in \Z[\vx]$ is called \emph{set-multilinear} if there exists a partition of $\vx$ into $d$ sets $\vx_1, \ldots, \vx_d$, and every monomial with non-zero coefficient in $f$ is of the form $\prod_{i=1}^d x_{i,j_i}$ with $x_{i,j_i} \in \vx_i$.

If we denote $n_i := |\vx_i|$ and $\vx_i = (x_{i,1},\ldots,x_{i,n_i})$, there is a natural correspondence between such polynomials and $d$-dimensional tensors of format $n_1 \times n_2 \times \cdots \times n_d$ (the $(i_1, \ldots, i_d)$ entry of the tensor is the coefficient of $x_{1,i_1} x_{2,i_2} \cdots x_{d,i_d}$).

A tensor of rank one corresponds to a set-multilinear polynomial of the form
\[ \ell_1(\vx_1) \cdot \ell_2(\vx_2) \cdots \ell_d(\vx_d),\] where for every $1 \le i \le d$, $\ell_i$ is a linear function in the variables of $\vx_i$.

\begin{definition}[Set Multilinear Formulas of Depth 3]
\label[definition]{def:set-mult-depth-3}
Let $f \in \Z[\vx]$ be a set-multilinear polynomial with respect to the partition $\vx=(\vx_1, \ldots, \vx_d)$. We say $f$ is computed by a set-multilinear formula of depth 3 and top fan-in $s$ over $\Z$ if
\[
f = \sum_{i=1}^s \prod_{j=1}^d \ell_{i,j}(\vx_j),
\]
where for every $1 \le i \le s$ and $1 \le j \le d$, $\ell_{i,j}$ is a linear function with integral coefficients.
\end{definition}

Set-multilinear polynomials computed by depth-3 formulas of top fan-in $s$ correspond to tensors of rank at most $s$.

We first prove that small depth-3 formulas compute Möbius-pseudorandom polynomials as in \cref{def:mob-polynomial-pseudorandom-eval}. Note that this proof does not even assume multilinearity or set-multilinearity of the formula.

\begin{theorem}
Let $f = \set{f_n}_{n \in \N}$ be a sequence of integer polynomials computed by depth-$3$ formulas of top fan-in $s$, where $s \le \varepsilon n$ for some small constant $\varepsilon > 0$. Then $f$ is Möbius-pseudorandom.
\end{theorem}

\begin{proof}
Let
\[
    f(\vx)=\sum_{i=1}^s P_i(x),
    \qquad
    P_i(\vx)=\prod_{j=1}^{d_i}\ell_{i,j}(\vx),
\]
where each \(\ell_{i,j}\) has integer coefficients.

Let $g_i(\vx)=(-1)^{P_i(\vx)}$
and $\chi_{i,j}(\vx)=(-1)^{\ell_{i,j}(\vx)}.$
Each \(\chi_{i,j}\) is a Fourier-Walsh character. Then, observe that
\begin{align*}
    g_i(\vx)
    &=
    1-2\prod_{j=1}^{d_i}
        \frac{1-\chi_{i,j}(\vx)}{2} \\
    &=
    1-2^{1-{d_i}}\prod_{j=1}^{d_i}
        (1-\chi_{i,j}(\vx)).
\end{align*}
Expanding the product and using the triangle inequality
gives
\[
    \alg{g_i}
    \leq
    1+2^{1-d_i}\cdot 2^{d_i}
    =3.
\]
Since $(-1)^{f(x)}=\prod_{i=1}^s g_i(x)$, submultiplicativity yields
\[
\alg{(-1)^f} \le \prod_{i=1}^s \alg{g_i}
    \leq 3^s.
\]
By \cref{lem:corr-algebra-norm}, assuming $s \le \varepsilon n$,
\[
\frac{1}{N} \left| \sum_{m=0}^{N-1} \mu(m) \exp(i \pi f(m_1, \ldots, m_n)) \right|  = o(1). \qedhere
\]
\end{proof}

\section{Polynomials with Low-Dimensional Partial Derivatives}
\label[section]{sec:pd}

The method of partial derivatives is a common technique for proving lower bounds for various computational models.

Let $f \in \F[\vx]$. Denote by $\partial f$ the space of partial derivatives (of all orders) of $f$. The following properties, by \cite{NW97}, are easy to prove:

\begin{claim}[\cite{NW97}]
\label[claim]{claim:pd-NW}
\begin{enumerate}
\item If $f = \ell^d$ where $\ell$ is a linear function, then $\dim \partial f \le d+1$.
\item If $f = \prod_{j=1}^d \ell_{j} (\vx_j)$ where $\ell_j$ is a linear function for every $1 \le j \le d$, then $\dim \partial f \le 2^d$.
\item $\dim \partial(f+g) \le \dim \partial f + \dim \partial g$.
\end{enumerate}
Therefore, if $f = \sum_{i=1}^s \ell_i^d$ then $\dim \partial f \le (d+1)s$, and if $f = \sum_{i=1}^s \prod_{j=1}^d \ell_{i,j} (\vx_j)$, it holds that $\dim \partial f \le s \cdot 2^d$.
\end{claim}

It follows that lower bounds on the dimension of the space of partial derivatives of $f$ can be used to prove lower bounds on the minimal size of a depth-$3$ powering formula or a depth-$3$ set-multilinear formula computing $f$. Such lower bounds are proved in \cite{NW97} for polynomials such as the elementary symmetric polynomial and the determinant.


We conjecture that any polynomial $f$ such that $\dim \partial f$ is small is Möbius-pseudorandom.
 
\begin{conjecture}
\label[conjecture]{conj:pd-mob-pr}
Suppose $f=\set{f_n}$ is a sequence of integral multilinear polynomials such that $\dim_\Q \partial f_n \le r$ for some ``small'' $r$.
  Then $f$ is Möbius-pseudorandom.
\end{conjecture}

When $r$ is very small, even a bound on the dimension of the first-order partial derivatives of $f$ can be used to prove \cref{conj:pd-mob-pr}.

\begin{theorem}
\label[theorem]{thm:pd-mob-pr}
Let $f = \set{f_n}_{n \in \N}$ be a sequence of integer polynomials and $r \le \varepsilon n$ for some small enough constant $\varepsilon>0$. Suppose that $\dim_{\Q} \set{\frac{\partial f_n}{\partial x_i} : 1 \le i \le n} \le r$. Then $f$ is Möbius-pseudorandom.
\end{theorem}

In order to prove \cref{thm:pd-mob-pr} we need the following lemma.

\begin{lemma}
\label[lemma]{lem:pd-essential-variables}
Suppose $f \in \Z[\vx]$ is a polynomial such that $\dim_{\Q} \set{\frac{\partial f}{\partial x_i} : 1 \le i \le n} = r $ for $r \le n$. Then, there exist $r$ linear functions with integral coefficients $\ell_1 (\vx), \ldots, \ell_r (\vx)$ and an $r$-variate polynomial $g$ with integer coefficients such that $f = g(\ell_1(\vx), \ldots, \ell_r(\vx))$.
\end{lemma}

If we only require $g$ and the $\ell_i$'s to have rational coefficients, this lemma is well known and can be easily proved using a change of basis and partial derivatives computation using the chain rule. Requiring that all coefficients in this expression are integers can be done using a similar proof, but involves a little more care. We will show that the change of basis can be done using a \emph{unimodular} matrix. Recall that a square integer matrix $U$ is unimodular if $\det U \in \set{-1,+1}$. The inverse of $U$ is then also integral.

\begin{claim}
\label[claim]{cl:unimodular}
Suppose $a_1, \ldots, a_n$ are integers such that $\gcd(a_1, \ldots, a_n)=1$. Then there exists a unimodular $n \times n$ matrix $U$ whose last column is $(a_1, \ldots, a_n)$.
\end{claim}

\begin{proof}
For two integers $u,v$, if $d=\gcd(u,v)$ and $p,q$ are the Bézout coefficients, that is, integers such that $pu + qv=d$, then note that
\[
\begin{bmatrix}
    p & q \\
    -v/d & u/d
\end{bmatrix}
\cdot
\begin{bmatrix}
    u \\ v
\end{bmatrix}
=
\begin{bmatrix}
    d \\ 0
\end{bmatrix},
\]
and the matrix on the left has determinant 1.
By repeatedly applying this transformation to 2 coordinates (leaving all other coordinates unchanged), starting from the vector $\va=(a_1, \ldots, a_n)$, we can make all coordinates zero except one coordinate (which we may permute to be the last coordinate). That coordinate would equal $\gcd(a_1, \ldots, a_n)=1$. Thus we found a unimodular matrix $U'$ such that $U'\va=e_n$ where $e_n$ is the $n$-th elementary basis vector. Letting $U=(U')^{-1}$, $\va=Ue_n$, so $\va$ is the last column of $U$. Note that $U$ is also unimodular. 
\end{proof}

\begin{proof}[Proof of \cref{lem:pd-essential-variables}]
We use induction on $n-r$. If $r=n$, there is nothing to prove. Suppose $r<n$. Then there are rational constants $c_1, \ldots, c_n$, not all zero, such that
\[
\sum_{i=1}^n c_i \frac{\partial f}{\partial x_i} = 0.
\]
By clearing denominators, we obtain \emph{integers} $a_1, \ldots, a_n$, not all zero, such that
\begin{equation}
\label{eq:pd-integer}
\sum_{i=1}^n a_i \frac{\partial f}{\partial x_i} = 0.
\end{equation}
Dividing by their gcd, we may further assume that $\gcd(a_1, \ldots, a_n)=1$. Let $U$ be a unimodular matrix as in \cref{cl:unimodular} and consider
\[
h(\vy) = f(U \vy),
\]
where $\vy=(y_1, \ldots, y_n)$.
Observe that $h \in \Z[\vy]$ since $f$ has integer coefficients and $U$ is integral. 
By the chain rule, and since the last column of $U$ is $\va$,
\[
\frac{\partial h}{\partial y_n} = 
\sum_{i=1}^n a_i \frac{\partial f}{\partial x_i}(U \vy) = 0.
\]
This means that $h$ does not depend on $y_n$ (here we use the assumption that we work over characteristic 0), so $h$ is in fact a polynomial in $n-1$ variables. Further, applying an invertible linear change of variables does not change the dimension of the first order partial derivatives, so the partial derivatives of $h$ still span a space whose dimension is $r$. By induction,
\[
h(y_1, \ldots, y_{n-1}) = g(L_1(\vy), \ldots, L_r(\vy))
\]
for some integral polynomial $g$ and integral linear functions $L_1, \ldots, L_r$, and therefore $\ell_i(\vx) = L_i (U^{-1} \vx)$ is also integral, and
\[
g(\ell_1(\vx), \ldots, \ell_r(\vx)) = g(L_1(U^{-1}\vx), \ldots, L_r(U^{-1}\vx)) = h(U^{-1}\vx) = f(\vx). \qedhere
\]
\end{proof}

We can now prove \cref{thm:pd-mob-pr}.

\begin{proof}[Proof of \cref{thm:pd-mob-pr}]
\sloppy
Let $\ell_1 (\vx), \ldots, \ell_r (\vx) \in \Z[\vx]$ and $g \in \Z[\vx]$ such that $f = g(\ell_1(\vx), \ldots, \ell_r(\vx))$, as in \cref{lem:pd-essential-variables}. By considering $f$ modulo 2, which involves reducing the coefficients of $g$ and $\ell_1, \ldots, \ell_r$ modulo 2, it follows that $(-1)^{f(\vx)}$ depends only on the values of $r$ linear functions modulo 2, so its Fourier-Walsh expansion involves at most $2^r$ non-zero coefficients. Therefore, by Cauchy-Schwarz, $\alg{(-1)^f} \le 2^{r/2}$. By \cref{lem:corr-algebra-norm} and by the assumption on $r$,
\[
\left| \frac{1}{N} \sum_{m=0}^{N-1} \mu(m) \exp(i \pi f(m_1, \ldots, m_n)) \right| =o(1)
\]
and the result follows.
\end{proof}

Note that since the dimension of first order partial derivatives of $f$ is at most $n$, \cref{thm:pd-mob-pr} is asymptotically optimal (although in general one might expect this dimension to be close to $n$, even for rather simple polynomials).

\subsection{Bounded Depth Algebraic Formulas}

An algebraic formula over a field $\F$  is a tree whose leaves are labeled by variables from the set $\set{x_1, \ldots, x_n}$ or constants from $\F$, and every internal node is labeled by either $+$ or $\times$. The formula computes a polynomial in the natural way, and the depth of the formula is the depth of the tree.

Far-reaching extensions of Nisan and Wigderson's basic idea \cite{NW97} were used for proving lower bounds on many algebraic models. Limaye, Srinivasan and Tavenas \cite{LST25} proved super-polynomial lower bounds for bounded-depth algebraic formulas using this technique (see also \cite{Forbes24} for an extension of this result for all fields). These recent works provide directions for extending \cref{conj:pd-mob-pr} in a way that will imply Liouville-pseudorandomness for bounded depth arithmetic circuits.

\begin{conjecture}
\label[conjecture]{conj:bounded-depth}
Suppose $f=\set{f_n}$ is a sequence of $n$-variate integer multilinear polynomials computed by depth-$\Delta$ formulas of size $s$, for $s=poly(n)$ and $\Delta=O(1)$. Then $f$ is Möbius-pseudorandom.
\end{conjecture}

\section{Circuits with Few Multiplication Gates}
\label[section]{sec:few-mult}

The results of the previous section extend to circuits with few multiplication gates.

\begin{theorem}
\label[theorem]{thm:few-multiplication-gates}
Suppose $f \in \Z[\vx]$ is computed by an integral arithmetic circuit with $k$ multiplication gates (with arbitrary fan-in). Then $\alg{(-1)^{f}} \le 3^k$.
\end{theorem}

\begin{proof}
Let $C$ be a circuit computing $f$. Reduce $C$ modulo $2$, and let $y_1, \ldots, y_k$ be the outputs of the $k$ multiplication gates, ordered in topological order. Observe that for every $i \in [k]$ we have that
\begin{equation}
\label{eq:y_i-as-function-of-smaller-ys}
y_i = \prod_{j=1}^{a_i} \ell_{i,j}(\vx,y_1, \ldots, y_{i-1})
\end{equation}
where $\ell_{i,j}$ is a linear function modulo $2$. Further, the output of the circuit, which is $f$ modulo 2, is an affine linear function in $\vx, y_1, \ldots, y_k$.

We first expand $(-1)^{f}$ as a function $f_k$ of $\vx, y_1, \ldots, y_k$. Observe that $f_k$ is a single Fourier-Walsh character with coefficient $\pm1$. 
We now eliminate $y_k, y_{k-1},\ldots,y_1$ in order to obtain the Fourier expansion of $(-1)^{f}$ as a function of $\vx$.

For $i=k, k-1, \ldots, 1$, suppose we have obtained a function $f_i$ that depends on $\vx$ and $y_{i},y_{i-1}, \ldots, y_1$. Whenever a character involves the variable $y_i$, we write it as $(-1)^{y_i} \cdot (-1)^{\ell'(\vx, y_1, \ldots, y_{i-1})}$ for some function $\ell'(\vx, y_1, \ldots, y_{i-1})$, and then replace $(-1)^{y_i}$ using the identity
\begin{equation}
\label{eq:yi-replacement}
(-1)^{y_i} = 1- 2 \prod_{j=1}^{a_i} \frac{1-(-1)^{\ell_{i,j}(\vx, y_1, \ldots, y_{i-1})}}{2}
\end{equation}
that follows from \eqref{eq:y_i-as-function-of-smaller-ys}.

The expression on the right hand side of \eqref{eq:yi-replacement} has Fourier-Walsh expansion whose algebra norm is at most $3$. Therefore, since the right hand side of \eqref{eq:yi-replacement} multiplies in the worst case every character of $f_i$, we have that after the replacement, $\alg{f_{i-1}} \le 3\alg{f_i}$.

Since $f_0=(-1)^f$ and $\alg{f_k}=1$, it follows that $\alg{(-1)^f} \le 3^k$.
\end{proof}

As an immediate corollary of \cref{thm:few-multiplication-gates} and \cref{lem:corr-algebra-norm}, we obtain the following.

\begin{corollary}
\label[corollary]{cor:few-multiplication-gates-pseudorandom}
There exists $\varepsilon>0$ such that for every $f \in \Z[\vx]$ computed by an integral arithmetic circuit with $k\le \varepsilon n$ multiplication gates (with arbitrary fan-in), it holds that $f$ is Möbius-pseudorandom.
\end{corollary}


\section{Möbius-Pseudorandomness of the Determinant}
\label[section]{sec:determinant}
Let $n:=k^2$ be a perfect square, and let $X=(x_{i,j})_{i,j \in [k]}$. The determinant polynomial is defined as $\Det_k = \sum_{\sigma \in S_k} (-1)^{\mathrm{sgn}(\sigma)}\prod_{i=1}^k x_{i,\sigma(i)} $. In this section we prove that $\Det_k$ is Möbius-Pseudorandom. Observe that since in \cref{def:mob-polynomial-pseudorandom-eval} we only look at the value of the determinant modulo 2, the same proof holds for the permanent.

The idea, as before, is to approximate the sign of the determinant by a function of small algebra norm, and then use \cref{lem:corr-algebra-norm}. 
The approximation we construct is exact when the kernel has small dimension, and matrices with larger kernels are sufficiently rare to control the approximation error.

\begin{lemma}
\label[lemma]{lem:det-approximation}
Let $\Det_k : \F_2^{k \times k} \to \set{0,1}$ as above. For every $1 \le t \le k$, there exists a function $D_t : \F_2^{k \times k} \to \RR$ such that $\alg{D_t} \le 2^{(k+1)t}$ and $\Ltwo{ \Det_k-D_t } = O(2^{-t/2})$.
\end{lemma}

\begin{proof}
For a matrix $X \in \F_2^{k \times k}$ denote by $R(X) = k-\rank(X) = \dim \ker X$.
Further, let $Z(X) = |\ker X|$.

For a fixed $v$, let $Z_v : \F_2^{k \times k} \to \set{0,1}$ be the function that equals $1$ when $Xv=0$ and zero otherwise.
We have that
\[
Z(X) = \sum_{v \in \F_2^k} Z_v(X)
\]
Further, the set $\{X : Xv=0\}$ is a linear space of matrices, so $Z_v$ is an indicator of a subspace. Therefore, $\alg{Z_v} = 1$ and $\alg{Z} \le 2^k$. Define now
\[
D_t(X) = \prod_{j=1}^t \frac{2^j - Z(X)}{2^j - 1}.
\]
$D_t$ is defined such that $D_t(X)=1$ whenever $Z(X)=1$ (which means that $X$ is non-singular, that is, $\Det_k(X)=1$), and $D_t(X) = 0$ whenever $Z(X) = 2^r$ for some $1 \le r \le t$, which means that $X$ has nullity at least $1$ and at most $t$ (and in particular, $\Det_k(X) = 0$).

By the submultiplicativity of the algebra norm,
\[
\alg{D_t} \le \prod_{j=1}^t \frac{ 2^j + \alg{Z}  }{2^j - 1} \le 2^{(k+1)t}.
\]
As argued above, $D_t(X) = \Det_k(X)$ whenever $R(X) \in \set{0, \ldots, t}$. If $R(X) = r > t$, we get that
\begin{align}
\label{eq:D_t}
|D_t(X)| &= \prod_{j=1}^t \frac{2^r - 2^j}{2^j - 1} \le \frac{2^{rt}}{\prod_{j=1}^t (2^j-1)} \nonumber \\
&=\frac{2^{rt - \binom{t+1}{2}}}{\prod_{j=1}^t (1-2^{-j})} = O\left(2^{rt-\binom{t+1}{2}}\right),
\end{align}
since $\prod_{j=1}^t (1-2^{-j})$ is bounded from below by an absolute constant.

Further, for a randomly chosen $X \in \F_2^{k \times k}$ we have that $\Pr[R(X) \ge r] \le O(2^{-r^2})$, as there are $O(2^{r(k-r)})$ subspaces of dimension $r$, each of which lies in $\ker X$ with probability $2^{-kr}$. Finally, we estimate
\[
\Ltwo{ \Det_k-D_t }^2 = \mathbb{E}_X[(\Det_k(X) - D_t(X))^2] 
\]
Conditioning on $R(X)=r$ for $r=0,\ldots,k$, the error is $0$ whenever $r \le t$ and, by \eqref{eq:D_t}, at most $O\left(2^{rt-\binom{t+1}{2}}\right)$ whenever $r \ge t$ (recall that $\Det_k(X)=0$ in this case), and therefore
\[
\Ltwo{ \Det_k-D_t }^2 =O\left(
\sum_{r=t+1}^k 2^{-r^2+2tr-t(t+1)} \right)
\]
Observing that $-r^2+2tr-t(t+1) = -t - (t-r)^2$, we get that
\[
\Ltwo{ \Det_k-D_t }^2 = O\left(
2^{-t} \sum_{s \ge 1} 2^{-s^2}
\right) = O(2^{-t}),
\]
as required by the lemma.
\end{proof}

\begin{theorem}
The $k \times k$ determinant polynomial on $n=k^2$ variables, $\Det_k$, is Möbius-pseudorandom.    
\end{theorem}

\begin{proof}
We first set parameters such that the function $D_t$ from \cref{lem:det-approximation} is Möbius-pseudorandom by virtue of \cref{lem:corr-algebra-norm}.
Indeed, set $t = k^{0.1}$ and let $D := D_t$ as in \cref{lem:det-approximation}. It then holds that $\alg{D} \le 2^{(k+1)t} = 2^{o(n)}$.

To apply \cref{lem:corr-algebra-norm}, we need a bound on $\Ltwo{D}$.
By writing $D = \Det_k + (D - \Det_k)$, it holds by \cref{lem:det-approximation} and the triangle inequality that 
\[
\Ltwo{ D } \le \Ltwo{\Det_k} + \Ltwo{D-\Det_k} \le 2
\]
for large enough $n$.

Since \cref{lem:corr-algebra-norm} as stated requires a function of $L_2$ norm at most $1$, we may define $\tilde{D} = D/2$, so that $\Ltwo{\tilde{D}} \le 1$ and $\alg{\tilde{D}} = 2^{o(n)}$.  By \cref{lem:corr-algebra-norm} we therefore get
\[
\left| \frac{1}{N} \sum_{m=0}^{N-1} \mu(m) \tilde{D}(m_1, \ldots, m_n) \right| = o(1),
\]
which clearly implies the same bound for $D$. 

We now move on to bound the correlation of $(-1)^{\Det_k}$ with the Möbius function. Since $\Det_k \bmod 2$ is $\set{0,1}$-valued,
\[
(-1)^{\Det_k} = 1-2\Det_k = 1 - 2(\Det_k - D) - 2D.
\]
Letting $E = \Det_k - D$, it therefore follows that
\begin{align*}
\left| \frac{1}{N} \sum_{m=0}^{N-1} \mu(m) (-1)^{\Det_k(m_1, \ldots, m_n)} \right| &\le
\left|\frac{1}{N} \sum_{m=0}^{N-1} \mu(m) \right| + 
2\left|\frac{1}{N} \sum_{m=0}^{N-1} \mu(m)E(m_1, \ldots, m_n) \right| \\
&+ 2 \left|\frac{1}{N} \sum_{m=0}^{N-1} \mu(m)D(m_1, \ldots, m_n) \right|.
\end{align*}
The first term above is $o(1)$ by the prime number theorem, and the third one is $o(1)$ by the argument above. As for the second one, by Cauchy-Schwarz we have that
\begin{align*}
\left|\frac{1}{N}\sum_{m=0}^{N-1} \mu(m)E(m_1, \ldots, m_n) \right|
& \le \left( \frac{1}{N} \sum_{m=0}^{N-1} \mu(m)^2 \right)^{1/2}
\left( \frac{1}{N}\sum_{m=0}^{N-1} E(m_1, \ldots, m_n)^2 \right)^{1/2} \\
&\le \left( \frac{1}{N}\sum_{m=0}^{N-1} E(m_1, \ldots, m_n)^2 \right)^{1/2} 
 = o(1)
\end{align*}
by
\cref{lem:det-approximation}.
\end{proof}

\subsection*{AI Use Disclosure}

The authors used ChatGPT Astra in all stages of this work, including for finding and verifying proofs, help in locating and understanding references, and proofreading.

\bibliographystyle{alphaurlpp}
\bibliography{references}

\end{document}